\documentclass{elsart}
\usepackage{epsfig}
\usepackage{graphicx}

\begin{document}

\begin{frontmatter}

\title{Studies of Avalanche Photodiode Performance in a High Magnetic Field}
\author{J. Marler, T. McCauley\thanksref{talk}},
\author{S. Reucroft, J. Swain}
\thanks[talk]{Corresponding author.
{\em E-mail address:\/} mccauley@eppserver.physics.neu.edu}
\address{Department of Physics, Northeastern University, Boston, Mass. 02115, USA}
\author{D. Budil, S. Kolaczkowski}
\address{Department of Chemistry, Northeastern University, Boston, Mass.
02115, USA}

\begin{abstract}
We report the results of exposing a Hamamatsu avalanche photodiode
(APD) to a 7.9 Tesla magnetic field. The effect of the magnetic
field on the gain of the APD is shown and discussed. We find APD
gain to be unaffected in the presence of such a magnetic field.
\end{abstract}

\begin{keyword}
Avalanche photodiode; Magnetic field
\end{keyword}

\end{frontmatter}

\section{Introduction}
The avalanche photodiode (APD) is a solid state photodiode with
internal gain. It has been chosen as the baseline photodetector
for the electromagnetic crystal calorimeter (ECAL) of the Compact
Muon Solenoid (CMS) Detector at the Large Hadron Collider at CERN
in Geneva, Switzerland. \cite{CMS} The ECAL consists of some
60,000 lead tungstate (PbWO$_{4}$) crystals, each to be read out
by a pair of APD's. Among the reasons for choosing the APD are
high quantum efficiency, a weak response to minimum ionizing
particles, otherwise known as the nuclear counter effect, and, as
we show here, an insensitivity to high magnetic fields. At the CMS
experiment, the magnetic field provided by the superconducting
solenoid will be 4 T. Here we present the results of tests carried
out to investigate the performance of a Hamamatsu APD, similar to
the one chosen for CMS, in the presence of a 7.9 T field. While it
has been widely believed that such a magnetic field would have
little or no effect on APD's, this is, as far as we know, the
first explicit measurement of APD performance in a strong magnetic
field with modern devices.

\section{APD Studied}
The APD studied was an experimental model developed by Hamamatsu
for CMS. The gain (M) of the APD, determined at bias voltage
V$_{\mathrm{b}}$, was calculated using Eq. 1. I$_{\mathrm{d}}$,
I$_{\mathrm{ill}}$, and I$_{\mathrm{ph}}$ are the dark current,
the illuminated current, and the photocurrent, respectively. The
photocurrent is the difference between the illuminated and dark
current. Illumination was provided by the light-emitting diode
(LED) described in Section 4.

\begin{equation}
M(V_{b})= \frac{I_{ill}(V_{b})-I_{d}(V_{b})}{I_{ill}(30
V)-I_{d}(30 V)}= \frac{I_{ph}(V_{b})}{I_{ph}(30 V)}
\end{equation}

The photocurrent at 30 V was used to determine the gain because in
this region of bias voltage, the photocurrent is essentially
constant and the gain is assumed to be equal to one. The working
bias voltage for a gain of 50 is approximately 355 V at a
temperature of 25 degrees C \cite{Yuri}. Other parameters
describing the APD are found in Table 1 \cite{Yuri}.

\begin{table}
\caption{Basic parameters of Hamamatsu APD}
\begin{tabular}{c c} \hline 
  Active area & $5 \times 5$ mm$^{2}$ \\
  Effective thickness (@M = 50) & 5.5 $\mu$m \\
  Capacitance & 125 pF \\
  Quantum efficiency (for $\lambda=480$ nm) & 76\% \\
  Excess noise factor (@M = 50) & 2.1 \\
  Temperature coefficient of gain (@M = 50) & 2.3\%/$^\circ$C \\
  \hline 
  \\
\end{tabular}
\end{table}

\section{The Magnetic Field}
The 7.9 T field was produced by a solenoidal magnet in the
High-Field EPR Laboratory at Northeastern University. The magnet,
produced by NMR Magnex Scientific Inc., features a horizontal
room-temperature bore design that allows convenient optical access
to the region of highest field homogeneity. The accessible region
of the magnet is a cylindrical volume of diameter 60.5 mm and
length 0.650 m. The field possesses homogeneity typical of magnets
used in NMR applications, with a maximum inhomogeneity of 0.2 ppm.
The region of maximum homogeneity is a cylindrical volume in the
center of 1 cm in diameter and 1 cm in length. Measurements were
performed with the APD in this cylindrical volume.

\section{APD Test}
\subsection{ Experimental Set-Up and Procedure}
The APD was mounted inside a light-tight container and was
connected to a blue light-emitting diode (LED) with optical fiber.
The blue LED was a model NSPB320BS produced by Nichia America
Corporation and emits light at a peak wavelength of 460 nm. The
set-up allowed for the surface of the APD to be oriented parallel
and perpendicular to the direction of the magnetic field. At each
of these orientations, the APD was inserted into the center of the
solenoid and the dark current (I$_{\mathrm{d}}$) and the
illuminated current (I$_{\mathrm{ill}}$) were measured as the bias
voltage (V$_{\mathrm{b}}$) was increased from 30 to 360 volts.
From these values, the gain (M) as a function of V$_{\mathrm{b}}$
was obtained using Eq. 1. For comparison, this procedure was also
repeated with the APD outside of the field. The bias source
voltage was provided by a Keithley 2410 Sourcemeter; the current
measurements were performed with this device as well.
\subsection{Analysis and Results}
The gain values for each value of the bias voltage for the runs
conducted outside of the field (five runs total) were averaged and
appear in Fig. 1. For each run conducted inside the field (one at
each orientation), and for each value of the bias voltage, the
gain was divided by the average value of the gain from the runs
conducted outside of the field. These values appear in Figs. 2-3.
The plots indicate that there appears to be no effect on the
performance of the APD in the presence of the magnetic field.

\begin{figure}
\begin{center}
\epsfig{file=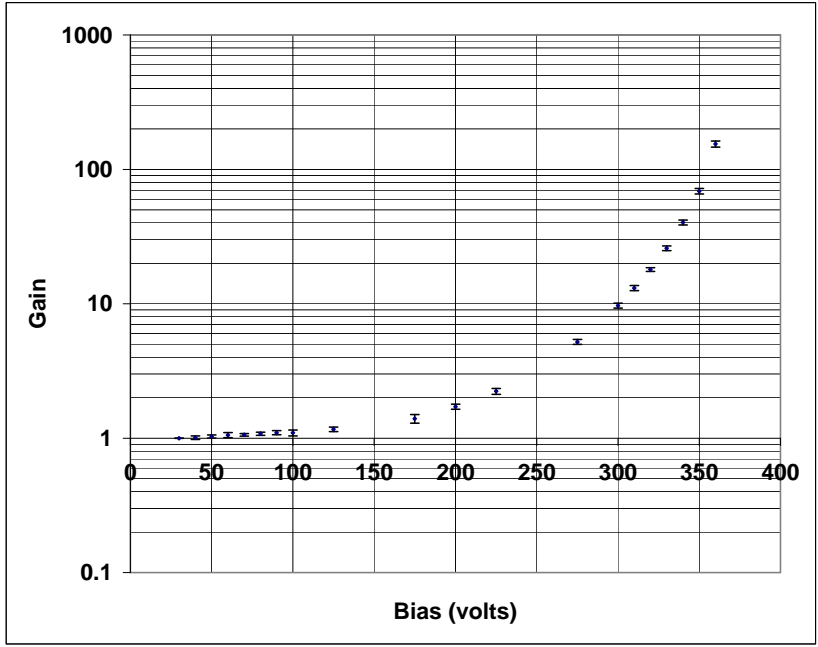,width=11cm,clip=} \caption {Average gain of
runs conducted outside the field vs. bias voltage.}
\end {center}
\end {figure}

\begin{figure}
\begin{center}
\epsfig{file=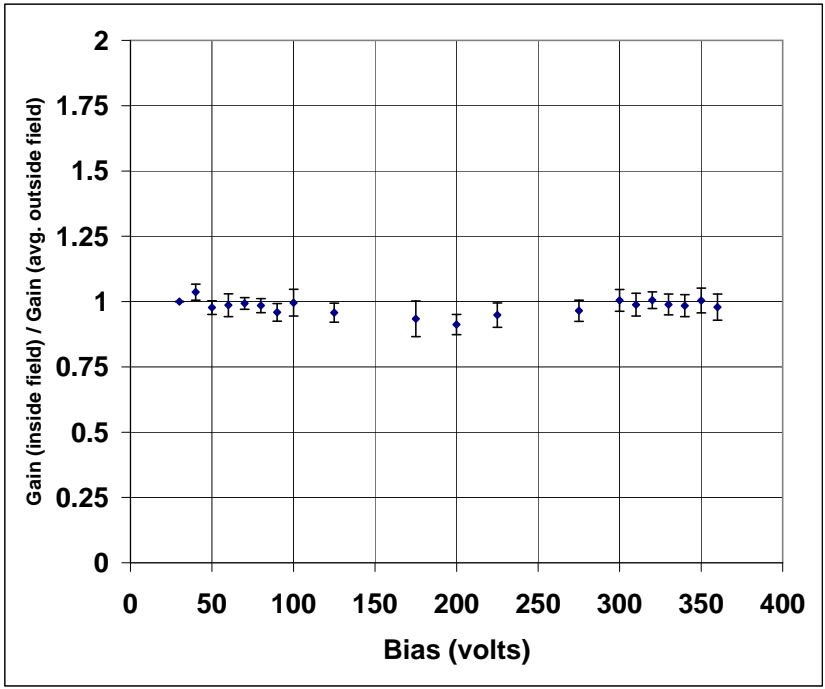,width=11cm,clip=} \caption {Comparison of
gain with surface of APD oriented perpendicular to the field with
average gain of runs conducted outside the field.}
\end {center}
\end {figure}

\begin{figure}
\begin{center}
\epsfig{file=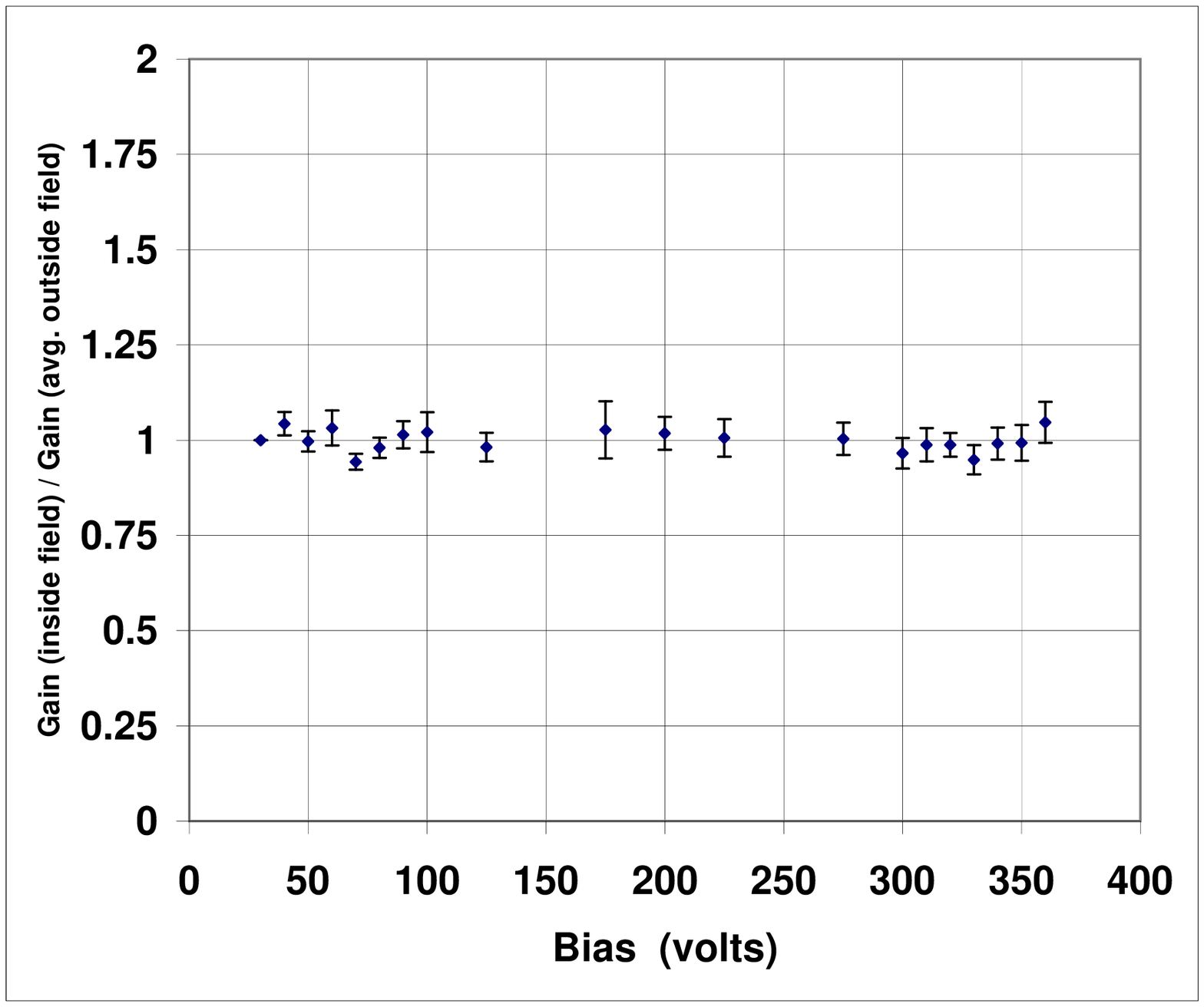,width=11cm,clip=} \caption {Comparison of
gain with surface of APD oriented parallel to the field with
average gain of runs conducted outside the field.}
\end {center}
\end {figure}

\section{Summary}
An experimental avalanche photodiode (APD) produced by Hamamatsu
was exposed to a 7.9 T magnetic field. The surface of the APD was
oriented both parallel and perpendicular to the field. At each
orientation, the dark current (I$_{\mathrm{d}}$) and illuminated
current (I$_{\mathrm{ill}}$) were measured as the bias voltage
(V$_{\mathrm{b}}$) was increased from 30 to 360 volts. From these
values the photocurrent (I$_{\mathrm{ph}}$) and gain for each
value of the bias voltage were obtained. For comparison, this
procedure was also performed with the APD outside of the magnetic
field. From this comparison, we find that APD gain is unaffected
by the presence of a 7.9 T magnetic field.

\ack We thank Y. Musienko for valuable advice and assistance, and
gratefully acknowledge the National Science Foundation for
financial support. We would also like to thank our colleagues on
CMS.


\begin{thebibliography}{9}

\bibitem{CMS}
{\em CMS Technical Proposal\/}, CERN, CERN/LHCC 94-38 LHCC/P1,
1994, {\em CMS ECAL Technical Design Report\/}, CERN, CERN/LHCC
97-33 CMS TDR 4, 1997.

\bibitem{Yuri}
Y. Musienko, private communication.

\end{thebibliography}
\end{document}